\documentclass[12pt]{article}
\usepackage{geometry}
\usepackage{graphicx}
\usepackage{amsmath, amssymb, amsthm, bm}
\usepackage[super]{natbib}
\usepackage{booktabs}
\usepackage{hyperref}
\usepackage{authblk}

\theoremstyle{definition}
\newtheorem{example}{Example}
\theoremstyle{plain}
\newtheorem{proposition}{Proposition}
\theoremstyle{definition}

\title{Hypothetical Treatment Accelerations: Estimating Causal Effects of Kidney Transplants from Observational Data}

\author[1]{Haris~Fawad}
\author[1]{Pål~Ryalen}
\author[2, 4, 5]{Vasiliki~Tsarpali}
\author[3, 4, 5]{Kristian~Heldal}
\author[1]{Kjetil~Røysland}
\affil[1]{Department of Biostatistics, University of Oslo}
\affil[2]{Institute of Clinical Medicine, University of Oslo}
\affil[3]{Institute of Health and Society, University of Oslo}
\affil[4]{Department of Transplantation Medicine, Oslo University Hospital}
\affil[5]{Clinic of Internal Medicine, Telemark Hospital Trust, Skien, Norway}

\date{}

\begin{document}
\maketitle

\begin{abstract}
Patients with end-stage kidney disease can expect to wait for several years before they receive a transplant, all the while their health deteriorates. How would the survival change if we managed to reduce these waiting times? To provide an answer, we present a continuous-time marginal structural model (MSM) of hypothetical scenarios in which the time until treatment is changed. In these scenarios, the treatment process is defined on a hypothetical time scale where time passes at a different rate compared to the time actually observed. Changing the time of the treatment process corresponds to changing the joint probability distribution, thereby making it possible to identify and estimate hypothetical parameters from observational data using previously developed methodology. We demonstrate this treatment-accelerated MSM using observational data from a Norwegian cohort of elderly patients with kidney failure. This model can potentially be useful to health authorities looking to assess the impacts of reducing the waiting times for organ transplantation in a given patient population. 
\end{abstract}


\section{Introduction}

Patients with end-stage kidney disease can expect to wait for several years before they receive a transplant. The median waiting time for a first kidney transplant ranges between 1.2 to 4.5 years internationally \cite{Wu2017globaltrendkidneyallocation}. It is the limited availability of organs that has created this bottleneck; there are not enough organs to ensure that every patient in need gets transplanted shortly after they enter the waiting list. We can imagine a hypothetical scenario in which the waiting times have been reduced, e.g. by an increase in the number of people who register as donors. 
Would the overall survival among patients with end-stage kidney disease change much in this scenario? 

To answer this question, we present a model for hypothetical scenarios in which the time until treatment is changed. This model for accelerating the treatment process is set in the framework of continuous-time marginal structural models (MSMs) \cite{roysland2022graphical, Roysland2011}. 

Mathematically, the treatment process is accelerated by composing it with a \textit{random change of time}. Time changes have been thoroughly studied in the continuous-time stochastic process literature, but have to the best of our knowledge not been used to model hypothetical scenarios. In simple terms, these time changes amount to speeding up the treatment process, or slowing it down, according to a specified rate factor. The assumption of causal validity \cite{roysland2022graphical} enables us to model these treatment accelerated scenarios using continuous-time MSMs. And the estimation strategy from \citet{Ryalen2019} ensures that we have consistent estimators for our model parameters. 

The main concepts in this paper are found in sections \ref{sec:change-of-time} and \ref{sec:change-of-measure}: Section \ref{sec:change-of-time} formalises treatment acceleration, and provides examples in the context of kidney transplantation; Section \ref{sec:change-of-measure} shows that a treatment acceleration corresponds to a change of probability measure, thereby making it possible to use previously developed methodology on continuous-time MSMs for inference. 
Next, Section \ref{sec:estimation} outlines the estimation of survival probabilities in the hypothetical treatment accelerated scenario using observational data; Section \ref{sec:kidney-transplant-analysis} presents a survival analysis of reducing the time to transplantation in a Norwegian cohort of elderly patients with end-stage kidney disease. Finally, Section \ref{sec:discussion} concludes with a discussion, and points towards extensions of treatment acceleration in the context of resource constraints.  

\section{Treatment accelerations}
\label{sec:change-of-time}

\subsection{Data structure}

We consider longitudinal data on $n$ i.i.d. patients over the study period $[0, \mathcal{T}]$. Each patient is represented by a vector of baseline variables $\mathcal{B}$ and a multivariate counting process $\mathcal{N} = (N_1, \dots, N_m)$. The counting processes include the treatment process $N_A$, which jumps from 0 to 1 at the time of treatment, and the outcome process $N_D$, which jumps from 0 to 1 at the time of the outcome event (e.g. death). Let $\{\mathcal F _t\}_t$ denote the history generated by $\mathcal B$ and the multivariate counting process $\mathcal{N}$.

\subsection{Time changes}

We want to model hypothetical scenarios in which the patients receive treatment at different times than at the observed times.

Let $\{g(t)\}_t$ be a strictly positive, left-continuous and $\mathcal F_t$-adapted process such that $\int_0^t \frac 1 {g(s)} ds < \infty$ for every $t > 0$, and let $\{\Gamma(t)\}_t$ be a process that solves the equation 
\begin{equation}
\Gamma(t) := \int_0^t g ( \Gamma(s) )ds. 
\label{eq:time change}
\end{equation}
Note that the process $\Gamma$ is strictly increasing since $g > 0$. We will model the desired hypothetical scenarios by requiring that a patient receives treatment $N_A(\Gamma(t))$ instead of $N_A(t)$. Before we can present this model, we need some notation. 
 
For an $\mathcal F_t$-adapted process $\{Z(t)\}_t$, we denote its time-shifted version by
\begin{equation*}
\check Z(t) := Z( \Gamma (t) ).
\end{equation*}
$\check Z$ may not be adapted to the original filtration $\{\mathcal F_t\}_t$. Therefore, we define another filtration $\{\check{\mathcal F}_t\}_t$, generated by $\mathcal B$ and the \textit{time-shifted} counting processes $\check N_1, \dots, \check N_m$. Then, since $Z$ is $\mathcal F_t$-adapted, $\check Z$ is $\check{\mathcal F}_t$-adapted. 

We will use continuous-time marginal structural models for counting processes \cite{Roysland2011} to model the hypothetical scenarios in which the time change $\Gamma$ has been applied to $N_A$. To that end, we need the intensity process of $\check N_A$.  
\begin{proposition}\label{proposition: treatment intensity on hypothetical time}
Let $P$ denote the probability measure that governs the frequencies of events in the observational data. If $\lambda_A$ is a $P$-intensity for $N_A$ with respect to the filtration $\{\mathcal F_t\}_t$ generated by the observational data, then
\begin{equation*}
\check g \cdot \check \lambda_A
\end{equation*}
defines a $P$-intensity for $\check N_A$ with respect to the filtration $\{\check {\mathcal F}_t\}_t$ generated by the time-shifted data. 
\end{proposition}

A proof is given in Appendix \ref{sec:appendix-time-transformation}.

In the next section we will use Proposition \ref{proposition: treatment intensity on hypothetical time} to identify model parameters (e.g. the hypothetical survival function). But first, let us look at some examples of hypothetical scenarios where $N_A$ is accelerated by $\Gamma$; these scenarios will be implemented in the analysis of the kidney transplant data in Section \ref{sec:kidney-transplant-analysis}.

\begin{example} \label{example: constant rate change}
Consider patients on a waiting list for kidney transplants. Clinical decision-makers may be interested in the effects of reducing the waiting times for transplants (on the survival of the population under study). We can specify a hypothetical scenario in which the waiting times have been changed by the same amount for all patients, by setting the slope of $\Gamma$ in \eqref{eq:time change} to:
\begin{equation*}
	\check g(t) = b.
\end{equation*}
That is, we accelerate $N_A$ by $\Gamma(t) = bt$. The choice $b > 1$ ($b < 1$) corresponds to a decrease (increase) in the waiting times. 
\end{example}

The time change $\Gamma$ is applied to an individual patient's treatment process; therefore, it can depend on a patient's covariates.

\begin{example}[Baseline covariate] \label{example: rate change depending on baseline}
Patients waiting for kidney transplants have baseline measurements of a comorbidity index ($X_{\text{LCI}}$). We can specify a hypothetical scenario in which the waiting times have been changed for patients with severe comorbidities (i.e. $X_{\text{LCI}} > 6$):
\begin{equation*}
	\check g(t, \mathcal{B}) = 1 + (b - 1) \cdot I(X_{\text{LCI}} > 6).
\end{equation*}
That is, we accelerate $N_A$ by $\Gamma(t) = bt$ for patients with severe comorbidities. The choice $b > 1$ ($b < 1$) corresponds to a decrease (increase) in the waiting times for these patients. In this hypothetical scenario, patients with moderate comorbidities (i.e. $X_{\text{LCI}} \leq 6$) receive the same treatment as in the observational data $N_A$ (since $\Gamma(t) = t$ in this sub-population).
\end{example}

\begin{example}[Time-varying covariate] \label{example: rate change with time-varying covariates}
The patients might be on dialysis while they wait for a kidney transplant.
Let $N_{\text{dialysis 2yr}}$ denote a counting process that jumps from 0 to 1 when a patient has been on dialysis for two years. We can specify a hypothetical scenario in which a patient's waiting time changes after they have been on dialysis for two years:
\begin{equation*}
	\check g(t, \mathcal{\check{N}}) = 1 + (b - 1) \cdot I(\check{N}_{\text{dialysis 2yr}}(t-) \neq 0).
\end{equation*}
That is, we accelerate $N_A$ by $\Gamma(t) = bt$ if the patient crosses the 2-year mark on dialysis.
Setting $b > 1$ ($b < 1$) corresponds to a decrease (increase) in the waiting times for these patients. As long as a patient as been on dialysis for less than two years, they receive the same treatment as in the observational data, since $\Gamma(t) = t$ during this time.
\end{example}

\section{Identification of the hypothetical scenario}
\label{sec:change-of-measure}

We want to model the frequencies of events we would have observed, had the patients received treatment $\check N_A$ instead of $N_A$. In that case, instead of observing the data $\mathcal B, N_1, \dots, N_m$, we would have observed the counterfactual data
\begin{equation*}
\mathcal B, \check N_A, N_k^{\check N_A} ; k \neq A .
\end{equation*}
(We assume that the underlying state space $\Omega$ is rich enough to realise such processes.) 

Obviously, we don't have access to such counterfactual data, but under under certain assumptions, we can still model their joint density: we can replace $P$ with another probability measure $P^g$ such that
\begin{equation*}
\text{Law}( \mathcal B, \check N_A,  N_k^{\check N_A}; k \neq A  | P ) = \text{Law}(\mathcal{B}, N_1, \dots, N_m | P^g).
\end{equation*}

Such laws on $\Omega$ are known to be determined from the density of $\mathcal B$ and from the intensities of the counting processes \cite{Jacod1975}. From Proposition \ref{proposition: treatment intensity on hypothetical time} we know that $\check{g} \check{\lambda}_A$ is a $P$-intensity for $\check N_A$, w.r.t. $\{\check{\mathcal{F}}_t\}_t$. This means that $g \lambda_A$, which is an $\mathcal{F}_t$-predictable process, gives the values of the intensity of $\check N_A$ for all $t$ (and for all $\omega \in \Omega$). To identify $P^g$, we also need the intensities of the remaining processes $ N_k; k \neq A$, and the density of $\mathcal B$. We will assume that these quantities do not change when we carry out the hypothetical intervention on $N_A$. This is called the assumption of \emph{causal validity} \cite{Ryalen2019, roysland2022graphical}.

\subsection{Causal validity}

The change of measure from $P$ to $P^g$ is said to be causally valid if $N_A$ has intensity process $g \lambda_A$ w.r.t. $P^g$, $N_k$ has intensity process $\lambda_k$ (for $k \neq A$) w.r.t. both $P$ and $P^g$, and $P(\mathcal{B}) = P^g(\mathcal{B})$.

This assumption amounts to there being no unmeasured confounders between the treatment and outcome processes ($N_A$ and $N_D$, respectively). Causal validity in systems with unmeasured variables/processes can be read off from causal local independence graphs using \textit{eliminability} \cite{roysland2022graphical}. We will discuss potential violations of causal validity by unmeasured variables/processes in the observational data on kidney transplants in Section \ref{sec:kidney-transplant-analysis}.

\subsection{Constructing $P^g$}

Both the observational measure $P$ and the hypothetical measure $P^g$ are uniquely characterized by the joint distribution of the variables in $\mathcal B$ and the predictable intensity processes of the counting processes in $\mathcal{N}$ \cite{Bremaud1981}. Moreover, if any event that has probability zero in the observational scenario also has probability zero in the hypothetical scenario, then $P^g$ can be obtained from $P$ by re-weighting according to the Radon-Nikodym derivative:
\begin{equation*}
    dP^g := R(\mathcal{T}) dP. 
\end{equation*}
Given causal validity, the process $\{R(t)\}_t$ (called the the likelihood ratio process) is a solution to the following stochastic differential equation \cite{Ryalen2019}: 
\begin{equation}
	R(t) = 1 + \int_0^t R(s-) ( g(s) - 1) dM_A(s),
	\label{eq:likelihood-ratio}
\end{equation}
with $\{M_A(t)\}_t$ a $P$-martingale, where $M_A(t) := N_A(t) - \int_0^t \lambda_A(s) ds$. Note that $R$ is determined by $g$, the process from \eqref{eq:time change} involved in accelerating $N_A$ by $\Gamma$.

The likelihood ratio $R$ is used to infer parameters in the hypothetical scenario described by $P^g$. In fact, the relation $P^g(B) = E_P\big[1_B \cdot R(t) \big]$ establishes $R(t)$ as a weight on the event $B \in \mathcal{F}_t$. Following \citet{Ryalen2019}, we will use estimates of $R$ as weights in the estimation procedure described in Section \ref{sec:estimation}, akin to IPTWs in discrete-time MSMs \cite{Robins2000b}.

\subsection{Estimand}

Our parameter of interest is the survival function in the hypothetical scenario described by $P^g$:
\begin{equation*}
	S^g(t) := P^g(T_D > t),
\end{equation*}
where $T_D$ is the time of the outcome event (e.g. death). 

We choose to focus on the survival function $S^g$ for the purpose of analysing the kidney transplant data in Section \ref{sec:kidney-transplant-analysis}. But in principle it is possible to investigate other parameters expressed by $P^g$ such as the restricted mean survival, cumulative incidence functions, etc.

\section{Estimation}
\label{sec:estimation}

We will now describe how to estimate the hypothetical parameter $S^g$ from observational data on $n$ i.i.d. individuals. An important step is to correctly specify a model for the observational treatment intensity $\lambda_A$.

\paragraph{Estimating the treatment intensity.}

We use an additive hazards model \cite{Aalen2008} to specify the observational treatment intensity for individual $i$:
\begin{equation}
	\lambda_{i, A}(t) = Y_{i, A}(t) \bm{L}_i^{\top}(t-) \bm{\beta}(t),
	\label{eq:additive-observational-strategy}
\end{equation}
where $Y_{i, A}$ is an at-risk indicator for treatment, and $\bm{\beta}$ is a vector of regression functions. It is crucial that the regressors $\bm{L}$ include all confounders between $N_A$ and $N_D$. We will illustrate this point in the context of the kidney transplant data in Section \ref{subsec: assumptions}.

The regression functions are estimated on their cumulative form $\bm{B}(t) = \int_0^t \bm{\beta}(s) ds$ by Aalen's additive hazard estimator $\hat {\bm B}$. In turn, we obtain the predicted values $d\hat{\Lambda}_{i, A}(t) := Y_{i, A}(t) \bm{L}_i^{\top}(t-) d\hat{\bm{B}}(t)$ and the martingale increments
$d\hat{M}_{i, A}(t) := Y_{i,A}(t) dN_{i, A}(t) - d\hat{\Lambda}_{i, A}(t)$.

\paragraph{Estimating the likelihood ratio.}

Both the regression model \eqref{eq:additive-observational-strategy} and the treatment acceleration $g_i$ determine the estimate of the likelihood ratio $R_i$. Specifically, $R_i$ is estimated by inserting $d\hat{M}_{i, A}$ into the integral equation \eqref{eq:likelihood-ratio}, along with an individual's specified treatment acceleration $g_i$:

\begin{equation}
    \hat{R}_i(t) = 1 + \int_0^t \hat{R}_i(s-) ( g_i(s) - 1) d\hat{M}_{i, A}(s).
    \label{eq:likelihood ratio estimate}
\end{equation}

\paragraph{Estimating the survival function.}

We estimate the survival function $S^g$ by first estimating the cumulative hazard $H^g(t) := \int_0^t h^g(s) ds$, where $h^g$ is the hazard rate of the outcome event in the hypothetical scenario described by $P^g$. The cumulative hazard $H^g$ is estimated by a weighted Nelson-Aalen estimator using the $\hat{R}_i$'s as weights:
\begin{equation}
	\hat{H}^g(t) = \int_0^t  \sum_{i=1}^n \frac{\hat{R}_i(s-) Y_{i, D}(s)}{\sum_{j=1}^n \hat{R}_j(s-) Y_{j, D}(s)} dN_{i, D}(s),
	\label{eq:Nelson-Aalen-weighted-D}
\end{equation}
where $Y_{i, D}$ is the at-risk indicator for the outcome event. This estimator is a special case of the additive hazard estimator in \citet[Sect. 2.2]{Ryalen2019}. 

Then, $S^g$ is estimated by a simple transformation of $\hat{H}^g$ \cite{Ryalen2018b}:
\begin{equation}
	\hat{S}^g(t) = 1 - \int_0^t \hat{S}^g(s-) d\hat{H}^g(s).
	\label{eq:survival-function-estimate}
\end{equation}
The estimators \eqref{eq:Nelson-Aalen-weighted-D} and \eqref{eq:survival-function-estimate} 
are consistent if \eqref{eq:additive-observational-strategy} is correctly specified as an additive hazards model under the assumption of independent censoring \cite{Ryalen2019}. To account for statistical uncertainty we propose using the bootstrap method.

\section{Causal effects of accelerating kidney transplants}
\label{sec:kidney-transplant-analysis}

We will now implement the hypothetical scenarios of Examples \ref{example: constant rate change}--\ref{example: rate change with time-varying covariates} from Section \ref{sec:change-of-time}. Our goal is to estimate the effects of reducing the waiting times for kidney transplants on the overall survival of the population under study.

\subsection{Observational data}

The data set consists of a Norwegian cohort of 251 elderly patients with end-stage kidney disease, wait-listed to receive transplants from deceased donors \cite{Tsarpali2021OlderKidneyPatients}. Table \ref{tab:kidney-cohort} shows descriptive statistics for this cohort. The patients were followed from the time they entered a waiting list for kidney transplants (earliest entry: Aug. 2010) until death (or administrative censoring at: Apr. 2021). Some patients were on dialysis before they entered the waiting list, still others were put on dialysis while on the waiting list. The majority of the patients received a transplant (73.3 \%). Those who received a transplant received it at different waiting times, likely at different levels of health deterioration. 

For some patients a clinical decision was made to withdraw them from the waiting list (17.1 \%) because of their poor health. We have chosen to combine withdrawal and death to form our composite endpoint. This choice can be justified by these patients' unfavorable survival prognosis, which makes withdrawal a proxy for later death.

\subsection{Assumptions}
\label{subsec: assumptions}

\begin{figure}
    \centering
     \includegraphics[width=0.6\textwidth]{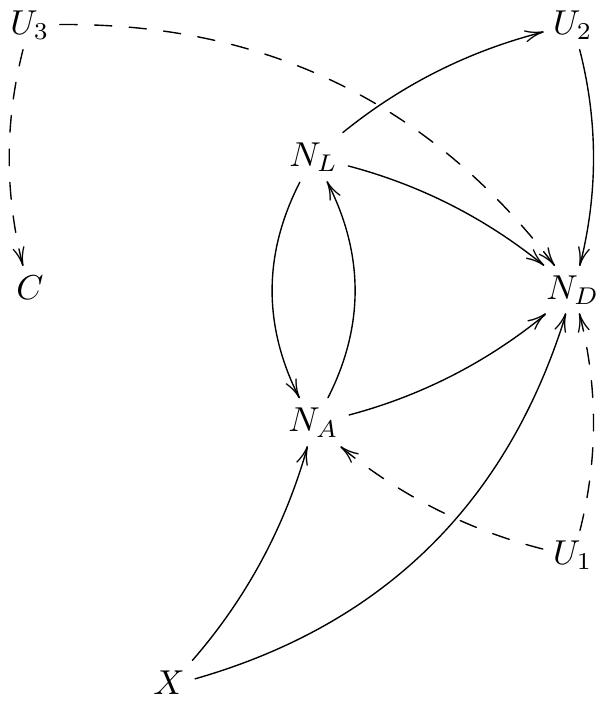} 
     \caption{A causal local independence graph of the kidney transplant data, where the intensity of the transplant process $N_A$ is intervened upon. The graph displays local independences for the processes stopped at the time of death/witdrawal $T_D$ \cite{Didelez2008}. The nodes $X$ and $N_L$ represent a baseline and a time-varying confounder, respectively; $C$ represents the censoring process (for death/witdrawal). $U_1, U_2,$ and $U_3$ represent unmeasured variables/processes; the remaining nodes are measured. The lack of an arrow between two nodes represents a local independence (we are assuming faithfulness of the model, to deduce from the graph if our identifying assumptions are violated). The dashed arrows indicate violations of our identifying assumptions: the presence of $U_3$ violates the independent censoring assumption for $N_D$, while the presence of $U_1$ violates causal validity as it is a common cause of both $N_A$ and $N_D$. However, causal validity is retained if either of the arrows out from $U_1$ is reversed. If we assume that the variables $U_1$ and $U_3$ are not present (and thus consider the subgraph induced by the nodes $X, N_L, N_A, N_D, C$ and $U_2$), then the presence of the unmeasured variable $U_2$ does not violate our identifying assumptions; $U_2$ is a so-called eliminable variable \cite{roysland2022graphical}.
     }
     \label{fig:obs-data-dependencies}
\end{figure}

The validity of our inferences relies on strong structural assumptions. We illustrate these assumptions in the local independence graph in Figure  \ref{fig:obs-data-dependencies}, in which dashed arrows show dependencies that violate our assumptions. Causal validity is violated by the presence of $U_1$, which represents any unmeasured common cause of receiving a transplant and experiencing death/withdrawal. Similarly, the presence of a common cause $U_3$ of receiving a transplant and being censored (for death/withdrawal) violates independent censoring. However, assuming that there are no such $U_1$ and $U_3$, the presence of $U_2$ (e.g. an unmeasured side effect of dialysis) does not violate our identifying assumptions. 

The assumptions we use are not the weakest possible assumptions that would ensure identification. For instance, we could allow for arrows from any measured node into the censoring node $C$. Lack of any such arrow reflects that we have ``staggered entry": each of the 147 censoring events are due to the time the data was extracted from the health registry (Apr. 2021), which is not caused by any variable in the graph. In the presence of the arrow $N_L \rightarrow C$ we would only have independent censoring conditioned on $N_L$, and we would then need to include censoring weights or impose a structural model for $N_D$ with $N_L$ as a covariate. 
Figure \ref{fig:obs-data-dependencies} shows that $X$ and $N_L$ are non-colliders that lie on so-called allowed paths from $N_A$ to $N_D$ \cite{roysland2022graphical}; thus, we need to include them in the regression model for $\lambda_A$ as specified below. 

\subsection{Observational treatment intensity}

In principle, organs are allocated with equity; patients who have waited the longest are most likely to receive a transplant. However, while the patients wait, they are likely to receive dialysis. Extended periods of dialysis is associated with decreasing health \cite{Heldal2009ClinicalOutcomesInElderly}, which may lead to a patient being taken off the waiting list due to his risk profile for transplantation. Other health related measurements that might act as confounders are: baseline comorbidities ($X_{\text{LCI}}$), any underlying disease that lead to a kidney failure diagnosis ($X_{\text{disease}}$), and a general physical function indicator ($N_{\text{physical}}(t)$, self-reported while on the waiting list) \cite{Tsarpali2022survival}.

Hence, the additive hazards model for the observed treatment intensity \eqref{eq:additive-observational-strategy} is specified as:
\begin{equation}
\begin{aligned}
	    \lambda_{i, A}(t) = Y_{i, A}(t) \{ &\beta_0(t) + \beta_1(t) \: I(X_{i, \text{LCI}} > 6) + \beta_2(t) \: X_{i, \text{disease}} + \\ &\beta_3(t) \: N_{i, \text{physical}}(t-) +\beta_4(t) \: I(N_{i, \text{dialysis 2yr}}(t-) \neq 0) \}.
	    \label{eq:observational-strategy}
\end{aligned}
\end{equation}
(This regression model appears to fit the data reasonably well, as shown by the martingale residuals in Figure \ref{fig:residuals} of Appendix \ref{sec:martingale residuals}.)

Using the predictive values of \eqref{eq:observational-strategy}, we can estimate $S^g$ under various treatment-accelerated scenarios by specifying $g_i$ in \eqref{eq:likelihood ratio estimate}.

\subsection{Accelerating transplants}

\begin{figure}
   \centering
   \includegraphics[width=1\textwidth]{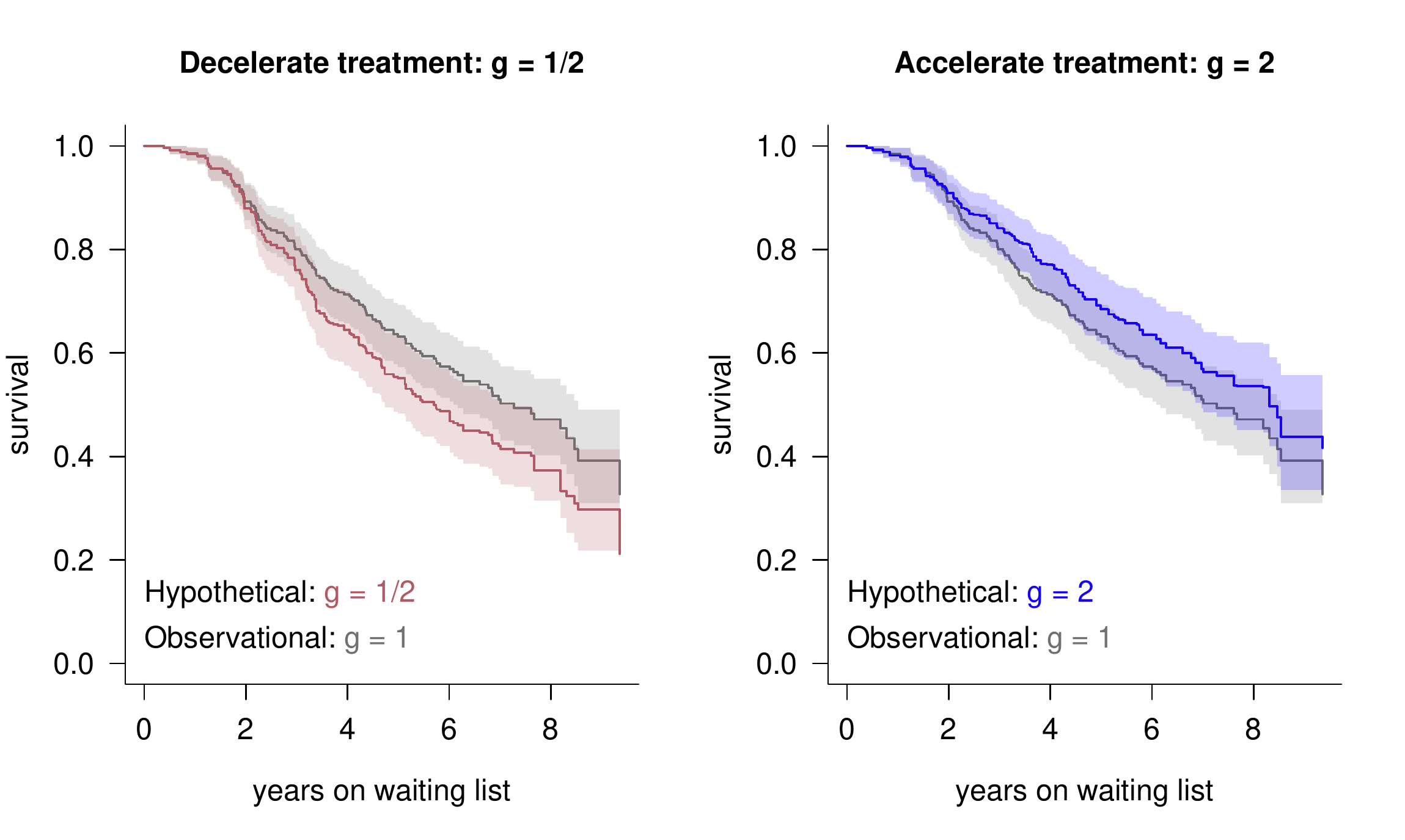} 
   \caption{Estimates of the survival function $\hat{S}^g$ of the composite end-point of death or withdrawal in two hypothetical scenarios --- defined by $g(t) = 1/2$ (red) and $g(t) = 2$ (blue) --- along with estimates of the corresponding survival function $\hat{S}$ in the observational data (gray). The shaded areas show 95 \% point-wise confidence intervals obtained by 1500 bootstrap iterations.}
   \label{fig:km-kidney-all}
\end{figure}

As in Example \ref{example: constant rate change}, let us accelerate the transplant process for an individual patient $N_{i, A}$ by $g(t) = b$, in effect changing the waiting times for all patients by a constant factor $b$. Figure \ref{fig:km-kidney-all} shows that when the transplant process is accelerated by $b = 2$ ($b = 1/2$), the estimated survival function $\hat{S}^g$ increases (decreases). In other words, survival increases as the waiting times are reduced. 

\subsection{Accelerating transplants in sub-populations}

Let us accelerate $N_{i, A}$ based on a patient's covariates, as in Examples \ref{example: rate change depending on baseline} and \ref{example: rate change with time-varying covariates}.

\begin{figure}
   \centering
   \includegraphics[height=0.75\textheight]{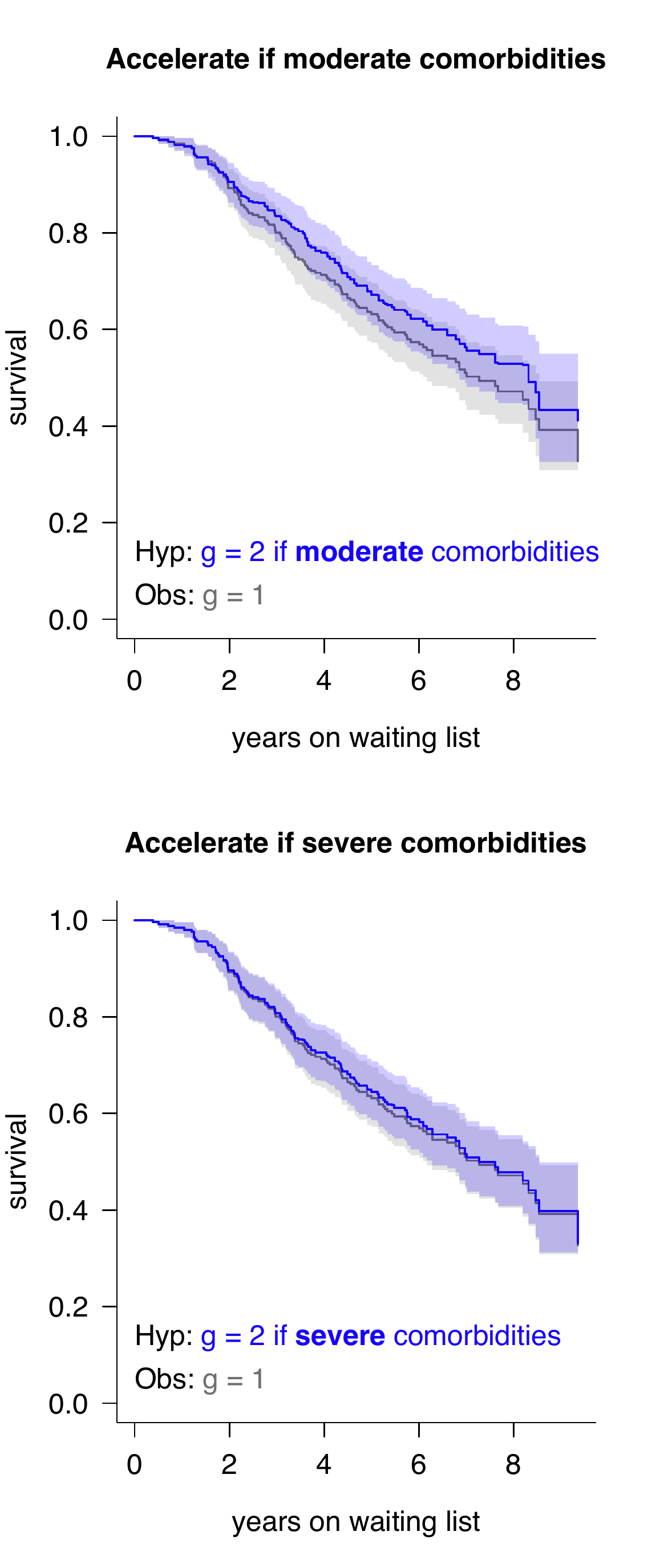} 
   \caption{Estimates of $\hat{S}^g$ in two hypothetical scenarios where transplants are accelerated (blue lines) based on a patient's baseline comorbidities. The top panel shows the result of accelerating patients with moderate comorbidities, by $g_1$ in  \eqref{eq:intervention on low comorbidity}; the bottom panel of those with severe comorbidities, by $g_2$ in \eqref{eq:intervention on high comorbidity}. Gray lines show the estimated survival $\hat{S}$ in the observational data ($g = 1$). The shaded areas show 95 \% point-wise confidence intervals obtained by 1500 bootstrap iterations.}
   \label{fig:km-kidney-liu}
\end{figure}

First, consider treatment accelerations based on baseline comorbidities ($X_{\text{LCI}}$). We can accelerate transplants for patients with moderate comorbidities:
\begin{equation}
    g_{1}(t, \mathcal{B}) = 1 + I(X_{i, \text{LCI}} \leq 6),
    \label{eq:intervention on low comorbidity}
\end{equation}
or for patients with severe comorbidities:
\begin{equation}
    g_{2}(t, \mathcal{B}) = 1 + I(X_{i, \text{LCI}} > 6).
    \label{eq:intervention on high comorbidity}
\end{equation}
(This latter acceleration is the same as in Example \ref{example: rate change depending on baseline}, with $b = 2$.) Figure \ref{fig:km-kidney-liu} shows that the estimated survival of the population $\hat{S}^g$ increases when the waiting times are reduced for patients with moderate comorbidities, but not so when they are reduced for patients with severe comorbidities. 

\begin{figure}
   \centering
   \includegraphics[height=0.75\textheight]{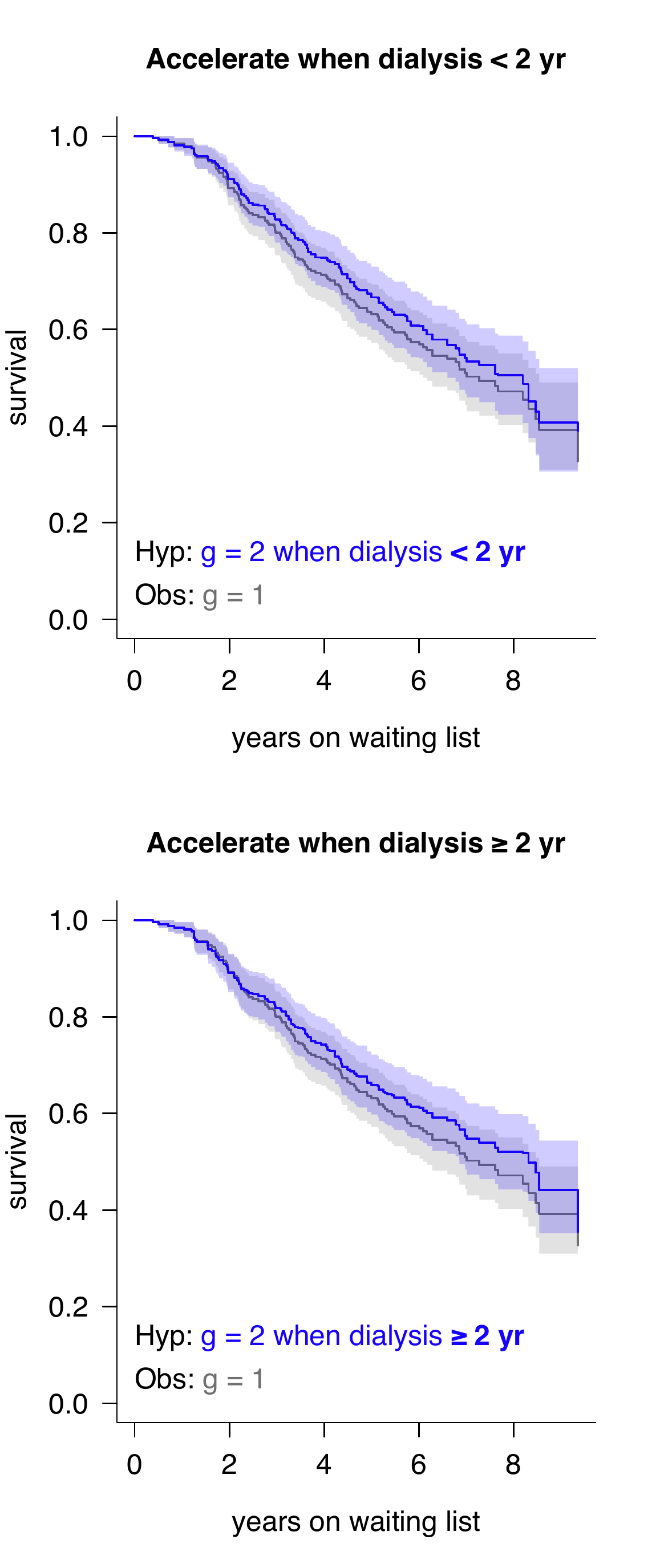} 
   \caption{Estimates of $\hat{S}^g$ in two hypothetical scenarios where transplants are accelerated (blue lines) based on a patient's time on dialysis. The top panel shows the result of accelerating patients who have been $< 2$ years on dialysis, by $g_3$ in \eqref{eq:intervention on dialysis less than 2}; the bottom panel of those who have been $\geq 2$ years on dialysis, by $g_4$ in \eqref{eq:intervention on dialysis more than 2}. Gray lines show the estimated survival $\hat{S}$ in the observational data ($g = 1$). The shaded areas show 95 \% point-wise confidence intervals obtained by 1500 bootstrap iterations.}
   \label{fig:km-kidney-dialysis}
\end{figure}

Next, consider accelerating transplants based on a patient's time on dialysis ($N_{\text{dialysis 2yr}}(t)$). We can accelerate the transplant process for patients who have been on dialysis for $< 2$ years:
\begin{equation}
    g_{3}(t, \mathcal{N}) = 1 + I(N_{i, \text{dialysis 2yr}}(t) = 0),
    \label{eq:intervention on dialysis less than 2}
\end{equation}
or for patients who have been on dialysis for $\geq 2$ years:
\begin{equation}
    g_{4}(t, \mathcal{N}) = 1 + I(N_{i, \text{dialysis 2yr}}(t) \neq 0).
    \label{eq:intervention on dialysis more than 2}
\end{equation}
(This latter acceleration is the same as in Example \ref{example: rate change with time-varying covariates}, with $b = 2$.) Figure \ref{fig:km-kidney-dialysis} shows that the survival of the study population increases in both cases.

\section{Discussion}
\label{sec:discussion}

The difference between the two treatment accelerations based on comorbidities, \eqref{eq:intervention on low comorbidity} and \eqref{eq:intervention on high comorbidity}, may lead us to believe that reducing the waiting times is less effective (with respect to the overall survival of the population) for patients with severe comorbidities than for patients with moderate comorbidities. However, the two hypothetical scenarios are not comparable for the purpose of a fair evaluation of reducing waiting times, even though each scenario is the result of accelerating a sub-population by a factor $g = 2$. This is because treatment acceleration models a relative change: by accelerating treatment according to $g$, the intensity process of treatment changes multiplicatively to $g \lambda_A$ (Proposition \ref{proposition: treatment intensity on hypothetical time}, and Section \ref{sec:change-of-measure}). In other words, although both $g_1 = 2$ and $g_2 = 2$ for their respective sub-populations, the observational treatment intensity $\lambda_A$ is likely to differ accross levels of comorbidities. This results in hypothetical scenarios with different waiting-time distributions.

The distribution of waiting times in the observational data can be thought of as a constraint when specifying hypothetical scenarios. This would amount to treatment accelerations which reduce the waiting times for some patients, and at the same time increase it for others, such that the overall distribution remains consistent with the observational data. We consider the characterisation of optimal waiting-time policies in continuous-time --- i.e. when there is a resource constraint on the treatment ---  a task for future work. Yet, the hypothetical scenarios presented in this paper might be of practical value for health authorities who would like to assess the effect of reducing the waiting times for transplantation when there \emph{is} an increase in the total availability of organs, e.g. by encouraging donor registrations, or expanding the criteria for transferable organs. 

In case researchers decide to model treatment accelerations on their data, they should address whether the assumption of causal validity holds. Using subject matter knowledge, they should list a sufficiently rich collection of baseline variables/processes, including those unobserved, to capture the important underlying mechanisms of the problem under study (as we did for the kidney transplant data in Section \ref{subsec: assumptions}). By encoding these variables/processes in a local independence graph, causal validity can be checked in the observed, possibly censored, data using eliminability. 

We provide a repository with implementation code at \url{https://github.com/harisf/hypotheticalTreatmentAccelerations}, that lets a user specify treatment accelerations $g$ on a simulated data set.

\section*{Funding}

This work was supported by the Research Council of Norway, grant number: 315323.

\clearpage
\bibliographystyle{unsrtnat}
\bibliography{references.bib}

\clearpage

\section{Figures \& tables}

\begin{table}
\centering
\caption{A Norwegian cohort of elderly patients with end-stage kidney disease, wait-listed to receive transplants from deceased donors. The original data set consisted of 289 patients from which we have excluded: 30 who received transplants from living donors, and 8 who had incomplete data on self-reported health variables --- resulting in 251 patients included for analysis.
}
\label{tab:kidney-cohort}
\begin{tabular}{l r r}
    \toprule
    Variable & Patients, $n = 251$ (100 \%) & Median (IQR) \\
    \midrule
    Age & & 70.6 (67.9--73.6) \\
    Event & & \\
    $\quad$ Withdrawn &  43 (17.1 \%) & \\
    $\quad$ Transplanted &  184 (73.3 \%) & \\
    $\quad$ Died &  & \\
    $\quad \quad$ Total & 104 (41.4 \%) & \\
    $\quad \quad$ Never transplanted & 45 (17.9 \%) \\
    $\quad$ Administratively censored &  147 (58.6 \%) & \\
    Liu comorbidity index $> 6$ & 28 (11.2 \%) & \\
    Cause for kidney disease &  & \\
     $\quad$ Diabetes & 24 (9.6 \%) & \\
     $\quad$ Vascular disease & 90 (35.9 \%) & \\
     $\quad$ Other & 137 (54.6 \%) & \\
    Dialysis-time* $\geq 2$ years & 42 (16.7 \%) & \\
    Self-reported physical function* & & \\
     $\quad$ at baseline & & 70.0 (47.5--80.0) \\
     $\quad$ at 6 months & & 70.0 (45.0--80.0) \\
     $\quad$ at 12 months & & 70.0 (41.3--80.0) \\
     $\quad$ at 18 months & & 60.0 (35.0--75.0) \\
    \bottomrule
\multicolumn{3}{l}{\small{*Before transplantation}}
\end{tabular}
\end{table}

\clearpage

\appendix

\section{Intensity of the accelerated treatment process}
\label{sec:appendix-time-transformation}

We want to show that 
\begin{equation}
\label{eq: paastand}
E_P[  \check{N}(\tau) ] = E_P\bigg[ \int_0^\tau (\check{g} \cdot \check{\lambda})(s) ds \bigg], 
\end{equation}
where $\tau$ is an $\check{\mathcal F}_t$-stopping time. 

Let $h$ be an $\mathcal{F}_t$-adapted bounded caglad process. Then
\begin{equation}
\label{eq:int telleprosess}
\int_0^t \check h(s) d\check N(s) = 
\sum_{s \leq t}  h( \Gamma(s)) \Delta N(\Gamma(s)) =
\sum_{u \leq \Gamma(t)}  h(u) \Delta N(u) = \int_0^{\Gamma(t)}  h(u) dN(u),
\end{equation}
using the variable change $u = \Gamma(s)$, and that $\Gamma$ is continuous and increasing.

Moreover, path-wise integration and a change of variables gives that
\begin{equation}
\label{eq:int intensitet}
\int_0^{\Gamma (t)} h(s) \lambda(s) ds = 
\int_0^{t} h( \Gamma (s)) \lambda( \Gamma(s) ) \Gamma'(s) ds = 
\int_0^{t} h( \Gamma (s)) \lambda( \Gamma(s) ) g(\Gamma(s)) ds, 
\end{equation}
where $\Gamma'$ denotes the left-hand derivative of $\Gamma$.

We will show that \eqref{eq:int telleprosess} and \eqref{eq:int intensitet} are equal in expectation by Doob's theorem. For that, we need to show that $\{ \Gamma (t) \}_t$ are stopping times w.r.t. $\{ \mathcal{F}_t \}_t$.

First, by elementary calculus, $\Gamma$ has path-wise inverse functions:
\begin{equation*}
\Gamma^{-1}(t) = 
\int_0^t \frac{1}{\Gamma'( \Gamma^{-1}(s) )} ds = 
\int_0^t \frac{1}{g(s)} ds. 
\end{equation*}
Then, since $g$ is $\mathcal{F}_t$-adapted
\begin{equation*}
\{ \Gamma(s) \leq t \} = 
\{ s \leq \Gamma^{-1}(t) \} = 
\bigg\{ s \leq \int_0^t \frac 1 {g(r)} dr \bigg\} \in \mathcal F_t
\end{equation*}
for every $s$ and $t$.

Indeed, since $h$ is $\mathcal{F}_t$-predictable, Doob's theorem gives that 
\begin{equation}
\label{eq:int likhet}
\begin{aligned}
E_P \bigg[ \int_0^t \check h(s) d\check N(s) \bigg] &= 
E_P \bigg[ \int_0^{\Gamma(t)} h(s) d N(s) \bigg] = 
E_P \bigg[ \int_0^{\Gamma(t)} h(s) \lambda(s) ds\bigg] \\ &= 
E_P \bigg[ \int_0^{t} \check h(s) \check \lambda( s) \check g(s)  ds \bigg].
\end{aligned}
\end{equation}

Now, suppose that $\tau$ is an $\check{\mathcal F}_t$-stopping time, and let $\tilde h(t) := I ( t \leq \tau )$. From \citet[Theorem I.6]{Protter2005}, we know that $\check {\mathcal {F}}_t = \mathcal F_{\Gamma(t)}$ (since $\Gamma (t)$ is an $\mathcal{F}_t$-stopping time). Since $\Gamma^{-1}$ is continuous, $h(t) := \tilde h(\Gamma^{-1}(t))$ defines an $\mathcal{F}_t$-adapted bounded caglad process. Inserting this $h$ into \eqref{eq:int likhet} proves our claim \eqref{eq: paastand}.

\section{Martingale residuals}
\label{sec:martingale residuals}

\begin{figure}
   \centering
   \includegraphics[width=0.9\textwidth]{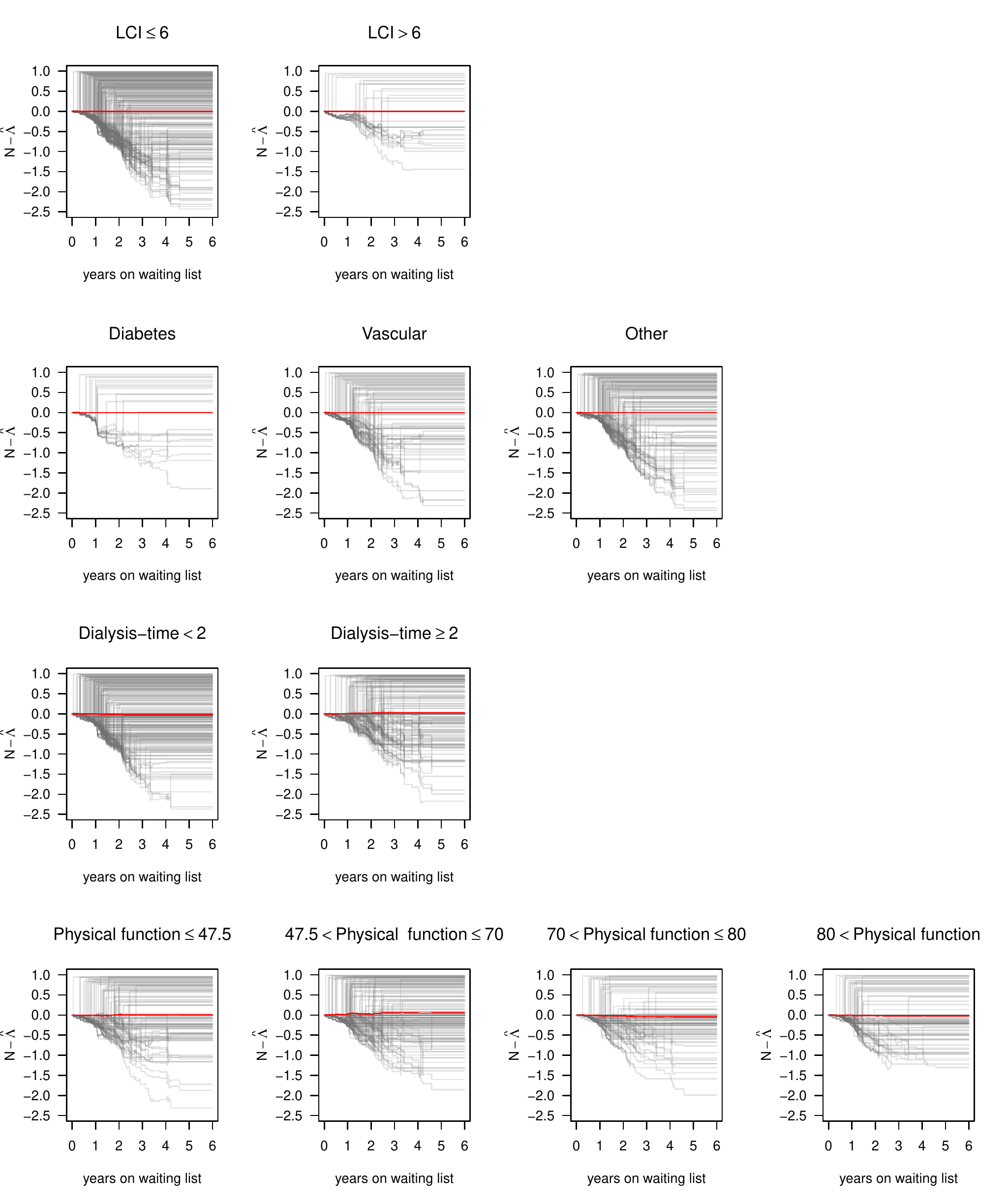} 
   \caption{Martingale residuals $\hat{M}_{i, A} = \int_0^t \{ Y_{i,A}(s) dN_{i, A}(s) - d\hat{\Lambda}_{i, A}(s) \}$ for each patient (gray), resulting from the regression model for the treatment intensity in \eqref{eq:observational-strategy}. Each row stratifies the patients according to the covariates in the model. The red line shows the average within each plot, which is expected to be close to zero.}
   \label{fig:residuals}
\end{figure}

\end{document}